\documentclass[prb,twocolumn,showpacs]{revtex4}
\pdfoutput=1
\usepackage{graphicx}
\usepackage{dcolumn}
\usepackage{amsmath}
\begin{document}

\title{Effects of impurities on Tamm-like lanthanide-metal surface states}

\author{D.\ Wegner}
\altaffiliation[Current address: ]{University of California at Berkeley, Department of Physics, Berkeley, CA 94720-7300}
\email{wegner@berkeley.edu}
\affiliation{Institut f{\"u}r Experimentalphysik, Freie Universit{\"a}t Berlin, Arnimallee 14, 14195 Berlin-Dahlem, Germany}
\author{A.\ Bauer}
\affiliation{Institut f{\"u}r Experimentalphysik, Freie Universit{\"a}t Berlin, Arnimallee 14, 14195 Berlin-Dahlem, Germany}
\author{G.\ Kaindl}
\affiliation{Institut f{\"u}r Experimentalphysik, Freie Universit{\"a}t Berlin, Arnimallee 14, 14195 Berlin-Dahlem, Germany}

\date{August 14, 2007}

\begin{abstract}
The effects of isolated residual-gas adsorbates on the local
electronic structure of the Dy(0001) surface were spatially mapped
by scanning tunneling microscopy and spectroscopy at 12~K. Less
than 15~{\AA} away from an adsorbate, a strong reduction of the
intensity and a significant increase of the width of the majority
component of the surface state due to impurity scattering were
observed, with essentially no change of  the minority component;
this reflects a high lateral localization of the Tamm-like surface
state. Furthermore, an adsorbate-induced state was found that
behaves metastable.
\end{abstract}


\maketitle

%
%

One challenge in surface science has always been the study of
pristine surfaces, and improvements in ultra-high vacuum
technology and surface preparation techniques have led to
continuous progress towards this goal. The absence of surface
impurities is particularly important when linewidths of surface
states, which are determined by hot-electron and hole dynamics,
i.e.\ by the lifetimes of excited electronic states, are
spectroscopically measured. \cite{Ech04} The role of clean
surfaces is best reflected in the development of spectroscopic
results on noble-metal surfaces over the recent decades.
\cite{Rei01} Problems due to surface impurities are particularly
severe for surface-integrating methods, like angle-resolved
photoelectron spectroscopy (ARPES), where it is almost impossible
to monitor the properties of pristine surfaces since even a single
adsorbate influences the electronic lifetime of the surface state
over distances of a few hundred {\AA}. \cite{Bur99} It has been
shown that laterally localized probes, such as scanning tunneling
microscopy (STM) and spectroscopy (STS), can reliably measure
linewidths on pristine parts of a surface. \cite{Kli00} Even
though later a careful ARPES study has been able to reproduce the
STS results on noble-metal surfaces, \cite{Nic00,Rei01} we
consider this an exception due to these rather inert metals. In
case of more reactive surfaces, such as those of the transition
metals, a comparable surface cleanness has not been reached so
far.

Recently, we demonstrated that STS is well suited for studying the
dynamics of excited surface states on lanthanide metals.
\cite{Bau02,Weg06b} The ability to identify and select locally
clean surface areas for spectroscopic studies is of utmost
importance for these highly reactive metals. However, a comparison
of linewidths on Gd(0001) determined by ARPES, \cite{Fed02} STS,
\cite{Reh03,Weg06b} and time-resolved photoemission \cite{Lou07}
calls for further studies regarding the role of defect-scattering
contributions on  the widths of surface states. \cite{Rei01,Bov07}
Therefore, we have studied the local influence of residual-gas
adsorbates on the electronic structure of the (0001) surface of
dysprosium metal (Dy) by STM and STS. We discuss the induced
changes in the majority and the minority components of the surface
state, and we report on the appearance of a novel
adsorbate-induced state. The lateral extensions of the
adsorbate-induced features provide quantitative information on the
localization of the Tamm-like Dy(0001) surface state with
$d_{z^2}$-symmetry. The results are compared with the properties
of the laterally more extended \emph{sp}-like Shockley-surface
states on noble metals.

%
%

The experiments were performed in ultrahigh vacuum (UHV, base
pressure $<3\times10^{-11}\, \text{mbar}$) with a home-built
low-temperature STM operated at 12~K. \cite{Bau02} A 30-monolayer
(ML) thick Dy film was deposited \emph{in situ} by electron-beam
evaporation of 99.99\% pure Dy metal from a Ta crucible onto a
clean W(110) single crystal kept at room temperature. With
thoroughly degassed evaporators, the vacuum stayed below
$5\times10^{-10}\, \text{mbar}$ during deposition. Subsequent
annealing of the film to $\simeq 800\, \text{K}$ led to a smooth
hcp(0001) surface. \cite{Bau02} STS spectra were measured with
fixed tip position and switched-off feedback control using
standard lock-in techniques (modulation amplitude 1~mV (rms),
modulation frequency $\simeq 360\, \text{Hz}$). As is well known,
the differential conductivity, $dI/dU$, is approximately
proportional to the local density of states (LDOS) of the surface.
Here, $I$ is the tunneling current and $U$ the sample bias
voltage. Typical acquisition times for a single spectrum ranged
from 10~seconds to 2~minutes. All spectra are raw data, i.e., the
mutual normalization depends only on the feedback parameters prior
to opening the feedback loop (here $-300\,\text{mV}$, 1~nA). The
sample bias was chosen such that tunneling into bulk states
dominates. The systematic error of this normalization can be
estimated to $\simeq 10\%$ by comparing the background intensities
(e.g.\ at $U<-200\,\text{mV}$).

%
%

\begin{figure}
\begin{center}
\includegraphics{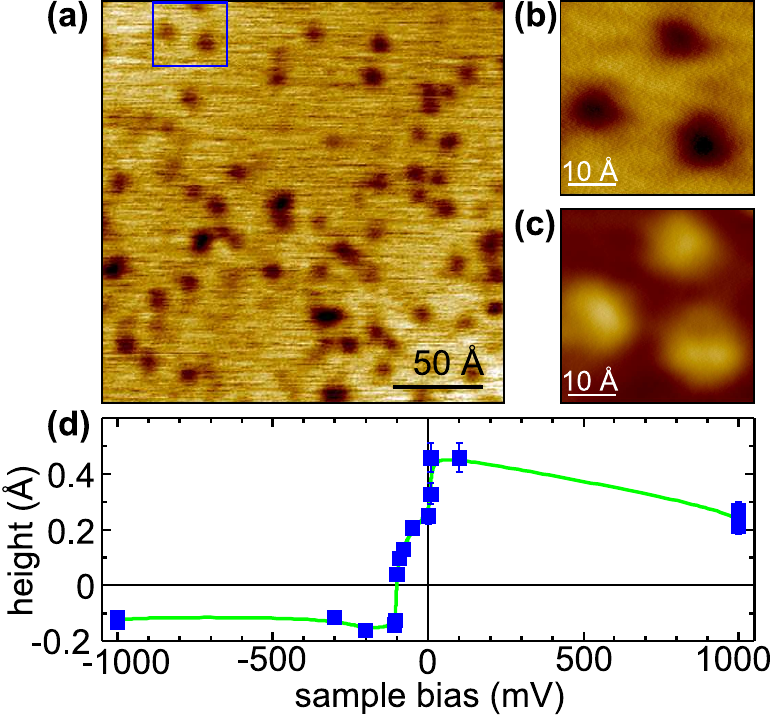}
\caption{\label{fig1} (Color online) (a)~STM topograph of a
Dy(0001) surface covered with $\simeq 0.01\,\text{ML}$ of hydrogen
adsorbates (see text; tunneling parameters: $-200\, \text{mV}$,
0.5~nA). A zoom of the marked area is shown at $-200\, \text{mV}$
(b) and $100\, \text{mV}$ (c). (d)~Bias dependence of the apparent
height of the adsorbates. A contrast reversal occurs at the
position of the MAC (see Fig.~\ref{fig2}).}
\end{center}
\end{figure}

Fig.~\ref{fig1} displays STM images of a terrace on the Dy(0001)
surface with residual-gas adsorbates of $\simeq 0.01\, \text{ML}$
nominal coverage. The apparent height of the adsorbates in the STM
images is bias dependent: Below $-100\, \text{mV}$, they appear as
depressions in the Dy film, while above $-100\, \text{mV}$, they
show up as protrusions. Even though we cannot identify with
certainty the nature of these adsorbates, there is strong evidence
that they are either atomic or molecular hydrogen, which is by far
the dominant impurity in well-degassed lanthanide-metal
evaporators. Furthermore, quadrupole mass spectroscopy during Dy
deposition revealed that the increase in pressure is almost
completely caused by a rising partial pressure of hydrogen. In
addition, hydrogen is known to cause a contrast reversal in STM
images of Gd(0001). \cite{Get99}

\begin{figure}
\begin{center}
\includegraphics{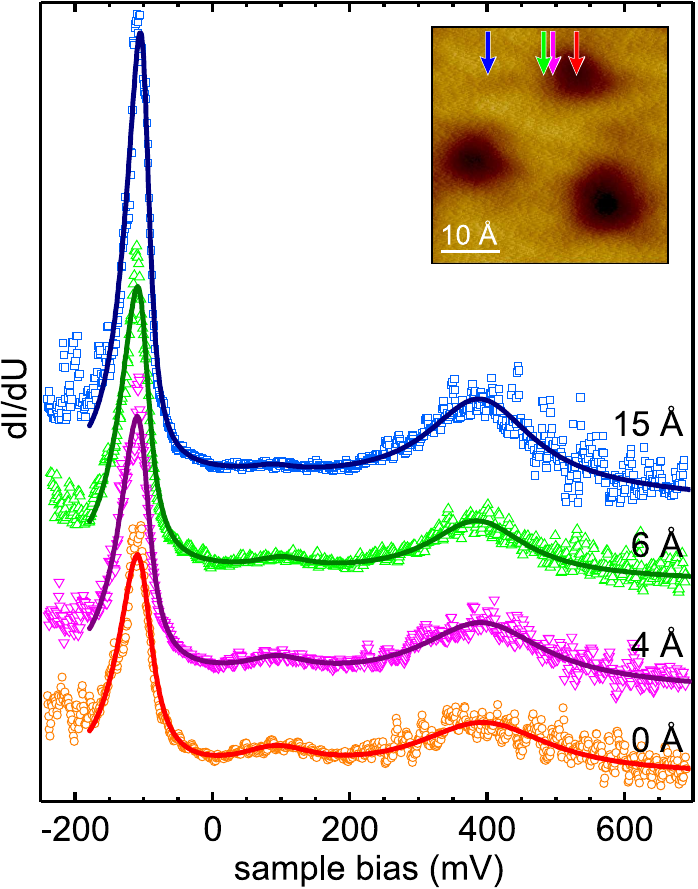}
\caption{\label{fig2}(Color online) Local tunneling spectra of a
Dy(0001) surface $\geq 15\, \text{\AA}$ away from any adsorbate
(top) and at various distances relative to an adsorbate (marked by
arrows in the STM image). The intensity of the MAC (at $\simeq
-100\, \text{mV}$) is halved on an adsorbate, while the width is
approximately doubled. In addition, an adsorbate-induced state
appears at $U \simeq 100\, \text{meV}$. The solid lines are
results of the fit analysis.}
\end{center}
\end{figure}

The observed bias dependence of the apparent height of the
adsorbates can be understood by comparing STS spectra on a clean
part of the Dy surface with those on adsorbates. Fig.~\ref{fig2}
displays representative STS spectra on an adsorbate (bottom
spectrum) and at different lateral distances away from an
adsorbate. The top spectrum (15 {\AA} away) resembles that of a
clean Dy(0001) surface, with a narrow peak at $\simeq -100\,
\text{mV}$ and a broad peak at $\simeq 400\, \text{mV}$, and also
their widths are equal to those observed for clean Dy surfaces.
These two peaks had previously been identified as the occupied
majority-spin component (MAC) and the unoccupied minority-spin
component (MIC) of the magnetically exchange-split surface state
with $d_{z^2}$ symmetry. \cite{Weg06b}

The presence of the adsorbate significantly alters the spectrum:
The LDOS of the MAC is halved, while its width is doubled. In
addition, a new peak appears that is only present close to an
adsorbate. This adsorbate-induced state can be described by a
thermally broadened Lorentzian with its maximum at $E_a = 97 \pm
8\, \text{meV}$ (i.e.\ it is unoccupied) and a width of $\Gamma_a
= 100 \pm 20\, \text{meV}$ (full width at half maximum, FWHM). The
width of the MIC remains essentially unchanged,\footnote{The
intrinsic scattering rate of the minority state is high, hence the
additional scattering channels due to the adsorbate seem to play
no major role.} and with the normalization described above, its
intensity is also approximately constant in all spectra (this
justifies the normalization additionally).

The apparent height in the STM images of these adsorbates vanishes
at a sample bias of $-100\, \text{mV}$ [see Fig.~\ref{fig1}(d)],
i.e., at the position of the MAC. The tunneling current $I$
corresponds to the integral of the LDOS between $E_F$ and the
applied bias voltage $U$. Since the LDOS of the MAC is reduced at
the adsorbate, the STM tip has to move closer to the surface for
$U < -100\, \text{mV}$ in order to keep $I$ constant, i.e., the
adsorbate appears as a depression. The apparent positive adsorbate
height for $U > -100\, \text{mV}$ is caused mainly by an
enhancement of the LDOS by the adsorbate-induced state (between 0
and 100 meV), and partly by a broadening of the MAC that actually
increases the LDOS for $U > -100\, \text{mV}$. Hence, the STM tip
has to retract away from the adsorbate in the constant-current
scanning mode.

In order to quantify the influence of adsorbates on the electronic
structure of Dy(0001), the series of spectra recorded at various
distances from adsorbates were fitted with a model previously
described. \cite{Bau02,Weg06} This model takes into account the
band dispersion of the surface state as well as thermal broadening
of the spectral features, and it yields the band maxima and the
linewidths (i.e.\ the inverse lifetimes) of the MAC and the MIC.
The adsorbate-induced state is described by a Lorentzian function.
The resulting fit curves are shown as solid lines in
Fig.~\ref{fig2}. It turns out that the only spectral features that
are systematically altered by the adsorbate are intensity and
width of the MAC as well as the intensity of the adsorbate-induced
state, while both position and width of the latter remain
unchanged. As discussed above, the MIC remains essentially
unchanged.

\begin{figure}
\begin{center}
\includegraphics{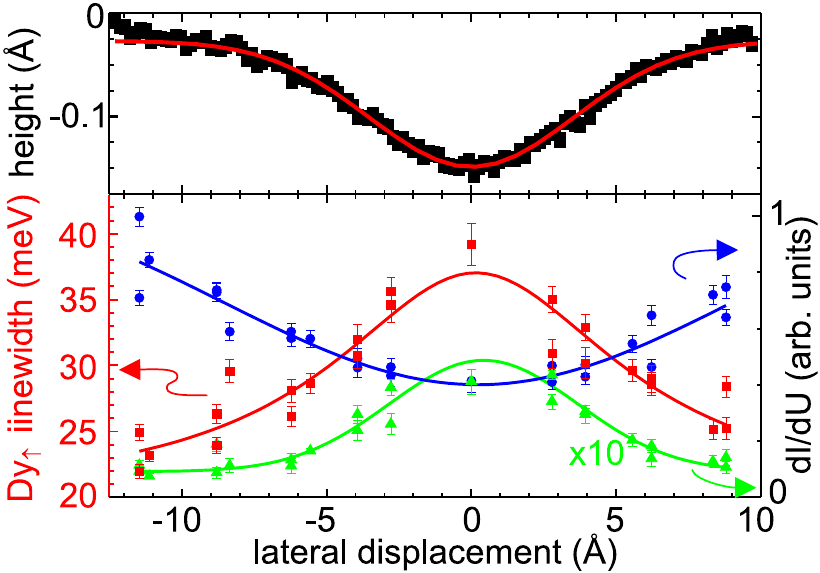}
\caption{\label{fig3}(Color online) Top: Topographical cross
section through a single adsorbate at $U = -300\, \text{mV}$.
Bottom: Lateral dependences of the width (Dy$_\uparrow$, red
squares) and the $dI/dU$ intensity of the MAC (blue circles) and
comparison with the $dI/dU$ intensity of the adsorbate-induced
state (green triangles).}
\end{center}
\end{figure}

Fig.~\ref{fig3} summarizes the lateral dependence of the fit
parameters, with the top panel displaying a linescan of the
topography through an adsorbate. A fit with a Gaussian results in
a width of $9 \pm 1\, \text{\AA}$ (FWHM). In the following, the
minimum of this topographical cross section is defined as the
center of the adsorbate. The intensity of the adsorbate-induced
state as a function of lateral displacement (green triangles in
the bottom panel) is also well described by a Gaussian, with its
maximum at the center of the adsorbate; its width is $8.3 \pm
0.5\, \text{\AA}$ (FWHM), identical to that of the topography. We
also find that $dI/dU$ of the MAC is reduced by $\approx 60\%$ at
the center of the adsorbates (blue circles). Note that the
increase of $dI/dU$ with increasing lateral distance from the
adsorbate occurs on a larger length scale, with a width of $17 \pm
4\, \text{\AA}$ (FWHM).

On the other hand, the lateral change of the linewidth of the MAC
(red squares) is again comparable to the behavior of the other
parameters, with a width of $12 \pm 2\, \text{\AA}$ (FWHM). At the
adsorbate, the linewidth is $\Gamma = 39 \pm 2\, \text{meV}$,
i.e., twice as large as the intrinsic linewidth on the pristine
surface ($\Gamma_i = 19 \pm 5\, \text{meV}$). \cite{Weg06b} Since
$\Gamma$ corresponds to the sum of the intrinsic and the
defect-scattering linewidth, the latter can be calculated as
$\Gamma_{\mathrm{def}} = \Gamma - \Gamma_i = 20 \pm 6\,
\text{meV}$; it shows that defect scattering doubles the inverse
lifetime of the MAC. We note that already at rather small
distances from the adsorbate ($> 15\, \text{\AA}$), the measured
linewidth equals the intrinsic linewidth of the defect-free
surface.

In summary, the spectroscopic features close to an adsorbate are
found to be influenced over lateral dimensions that correspond to
the extension of the adsorbate as reflected in the topographical
cross section (see Fig.~\ref{fig3}, top panel), i.e.\ $\simeq 10\,
\text{\AA}$. These dimensions reflect the localization of the
surface state and are in conformity with the very flat dispersions
of the surface states on these closed-packed lanthanide-metal
surfaces corresponding to large effective masses $|m^*|/m > 5$.
\cite{Bau02,Weg06b} If we neglect the contribution of the
adsorbate and ascribe the observed lateral changes to the surface
state alone, we arrive at the conclusion that the overlap of
$d_{z^2}$-like orbitals of two surface atoms at a distance larger
than three lattice constants (with $a = 3.59\, \text{\AA}$) must
be negligibly small. This supports a recent finding that the
exchange splittings of lanthanide-metal surface states deviate
from the overall bulk magnetizations. \cite{Bod99,Weg06c} Due to
high spatial localization, these Tamm-like surface states are more
sensitive to short-range than to long-range magnetic order.

Most of the described observations can be generalized to the
series of lanthanide metals as a whole. While the
adsorbate-induced state was observed solely on Dy(0001), all the
other lanthanide metals that have been studied so far (cf.\
Ref.~\onlinecite{Weg06b}) exhibit analogous surface defects with
comparable effects on the MAC. On the (0001) surfaces of Gd, Tb,
and Er, e.g., adsorbates were found that exhibit contrast
reversals in the apparent heights of STM images at bias voltages
of $-200\, \text{mV}$, $-100\, \text{mV}$, and $-50\, \text{mV}$,
respectively. In all three cases, these voltages correspond to the
positions of the respective MACs, and the contrast change can
again be explained by a reduction of the surface-state intensity
and a simultaneous broadening of the peak. On Lu(0001), adsorbates
were identified that also reduce and broaden the surface state.
However, a contrast reversal could not be observed in this case,
presumably due to the fact that the surface state in this case is
directly at $E_F$. \cite{Weg06} In all cases, the lateral
extensions of the adsorbates were found to be between 9 and
12~{\AA}.

A comparison with Shockley-type surface states on noble-metal
surfaces reveals that the lateral extensions of these states are
an order of magnitude larger than those of Tamm-like surface
states on lanthanide-metal surfaces. The explanation of this
significant difference is simple: The lifetimes, $\tau$, of
surface states on lanthanide metals are much shorter than those on
noble metals. Furthermore, the group velocities, $v_g$, of
electrons or holes in surface bands are smaller for
lanthanide-metal surfaces than for noble-metal surfaces due to
larger effective masses. Since the mean-free path is the product
of these two quantities, i.e., $\lambda = v_g \tau$, it is
dramatically reduced in case of the lanthanide metals. Whereas
electrons and holes in noble-metal surface states are considered
to be good approximations for a quasi-free electron gas, the
present results make it clear that this picture breaks down for
the surface states on lanthanide metals. Their small $\lambda$
also explains why STM images of lanthanide-metal surfaces do not
exhibit standing-wave patterns typical for the (111) surfaces of
noble metals.

The strong effects of an adsorbate on the linewidth of the MAC
explain why in case of Gd(0001) even high-resolution ARPES
measurements led to much larger linewidths than STS,
\cite{Fed02,Reh03} since the latter studies can be performed on
locally defect-free surface areas. Due to the high localization of
the surface state, such a measurement equals that on a defect-free
surface. Since the smallest spot probed by ARPES has typically a
diameter of $\simeq 1\, \mu\text{m}$, and the reactivity of the
lanthanide metals does not allow a defect-free area of this size,
we expect ARPES to measure an average width of clean and
adsorbate-influenced surface states. In case of Gd(0001), the STS
linewidth is about $2/3$ of that measured by ARPES. \cite{Reh03}

\begin{figure}
\begin{center}
\includegraphics{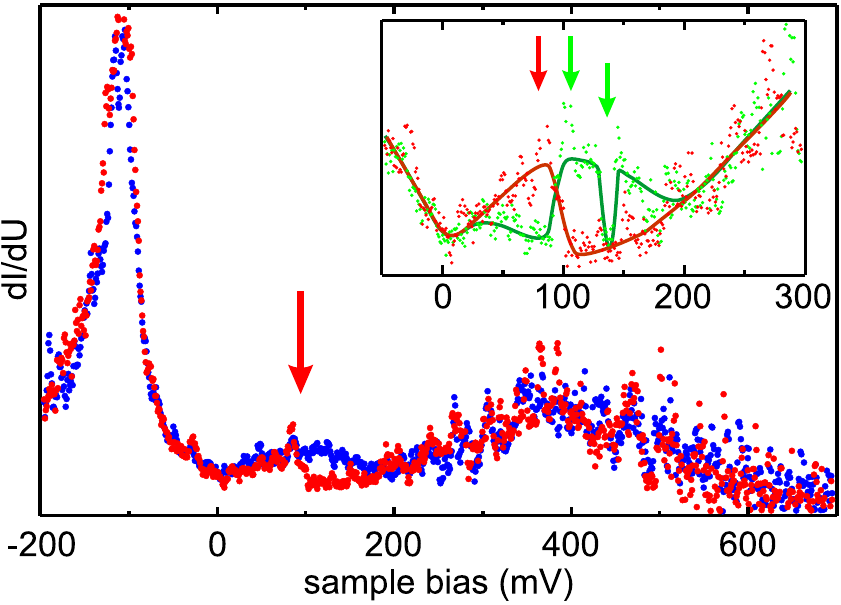}
\caption{\label{fig4}(Color online) Some of the STS spectra
exhibit signal jumps at a bias voltage corresponding to the
position of the adsorbate-induced state ($U \simeq 100\,
\text{mV}$; see arrow in red/grey spectrum), while others do not
(blue/black spectrum). Inset: Jumps can also occur "upwards" at
$\simeq 100\, \text{mV}$, as if the state suddenly appears at this
bias (see green/grey spectrum), and even several jumps can occur
during data acquisition.}
\end{center}
\end{figure}

Finally, we discuss an interesting detail in the spectroscopy of
the adsorbate-induced state: For several STS spectra, the $dI/dU$
signal suddenly jumps to zero during STS data acquisition, as if
the state had disappeared (Fig.~\ref{fig4}). These jumps occur
preferentially close to the peak of the adsorbate-induced state.
Also upward jumps occurred indicating the recovery of the
adsorbate-induced state, and even multiple jumps could be observed
during data acquisition (see inset to Fig.~\ref{fig4}).

Even though the origin for these jumps is unknown, it is obvious
to assume that they are caused by the influence of the tunneling
current on the adsorbate state itself, e.g.\ by electronically or
vibronically exciting the adsorbate or by charging it. Then, no
further electrons can tunnel into this state, and the LDOS feature
of the adsorbate-induced state disappears in STS. As the jumps
occur preferentially around 100~mV and no dependence on the
tip-sample distance was observed, we infer that a tunneling
electron must have sufficient energy to switch the adsorbate to an
excited state. Usually, each jump-down of the adsorbate state
(i.e., when it disappears in the spectrum) is followed by an
upward jump (i.e., when the state reappears); this indicates that
the excited state is metastable: within the timescale of a few
seconds, the adsorbate relaxes to its ground state. Consistently,
STM images at a bias of 100~mV have a three times larger noise
along the slow scanning direction with respect to images taken
below 10~mV or above 1~V. \footnote{Along the fast scanning
direction, no influence was observed.}

Further studies are necessary to clarify the exact nature of the
adsorbates (atomic or molecular hydrogen), e.g.\ by experiments
with controlled deposition of various gases on clean Dy(0001)
surfaces as well as by experiments aiming to clarify the nature of
the switching process. It will be interesting to check whether
this switching behavior can be influenced or controlled in a
reliable way. Note that the lifetimes of these excitations seem to
be of the order of 1 to 10~seconds, which is unusually long.

In summary, we have shown that single residual-gas impurities
(most probably hydrogen) on Dy(0001) exert strong influences on
the LDOS as well as on the linewidths of the occupied MAC. In
addition, a metastable adsorbate-induced state appears in the STS
spectra. All of these features are confined within $10\,
\text{\AA}$ (Gaussian FWHM) around the adsorbate, which is a
direct measure of the spatial extension of the surface state
itself. This confirms recent studies of lanthanide-metal surfaces,
which could only be understood by assuming that the surface states
are fairly localized.

%
%
The work was supported by the Deutsche Forschungsgemeinschaft
(DFG), projects Sfb-290/TPA6 and KA 564/10-1. A.B.\ acknowledges
support within the Heisenberg program of the DFG. D.W.\ is
grateful for funding by the Alexander von Humboldt Foundation.

\bibliography{defect}

\end{document}